# Unconventional Photon blockade in a Photonic Molecule Containing a Quantum Dot

Xiang Cheng, Han Ye, Zhongyuan Yu*

*State Key Laboratory of Information Photonics and Optical Communications, Beijing University of Posts and Telecommunications, P.O. Box 72, Beijing 100876, China*

*yuzhongyuan30@hotmail.com

Abstract: We propose a scheme to realize strong photon antibunching with lower photon nonlinearity in a photonic molecule consisting of two photonic cavities, one of which contains a quantum dot (QD). This strong photon antibunching is attributed to destructive quantum interference effect which suppresses the two-photon excitation of the cavity field. That $g^2(0) \approx 10^{-4}$ can be achieved with modest QD-cavity coupling strength $g = 1.1\kappa$ and cavity tunneling strength $J = 3\kappa$ when the system is driven by single laser field. To further reduce the requisite tunneling and make the system tunable, two laser fields are applied to the system. The strong photon antibunching ( $g^2(0) \approx 10^{-3}$) can be achieved with a relatively large intracavity photon number by optimizing the phase between two driving laser fields when $J = 0.9\kappa$. Moreover, the system shows a strong robustness of maintaining strong photon antibunching within a large parameter variation under the optimal phase condition. Our scheme provides a flexible and efficient method for solid state quantum single photon sources.

## 1. Introduction

Single photon source plays an important role in quantum information science for its significant applications in quantum cryptography [1], quantum key distribution [2-3], and linear optics quantum information processing [4]. Ideal single photon source emits photon singly thus the photons exhibit antibunching characteristics. Photon blockade effect that the first photon generated in cavity blocks the transmission of the second photon due to the nonlinearity of unequal level spacing in the J-C ladder, is one of the mechanisms to generate antibunching photons [5-7]. Lots of experiments have been implemented to realize photon blockade with a quantum dot (QD) in a microcavity [8-9], an atom in Rydberg state [10], circuit quantum electrodynamics system [11-12], etc. However, for a solid-state QD-cavity system, the strong coupling condition with $g \gg \kappa$ is hard to achieve on account of the challenges of nowadays fabrication techniques for high-quality microcavities [13-15], where *g* is the QD-cavity coupling strength and $\kappa$ is the cavity decay rate. To avoid the fabrication challenges, the photon blockade with strong sub-Poissonian light statistics based on bimodal-cavity system has been theoretically proposed [13, 16]. But, strong photon nonlinearity is not easy to achieve at single photon level for many systems, and the intracavity photon number is also low under strong photon blockade regime. A coupled single QD-cavity system was proposed to realize strong photon blockade utilizing quantum interference with lower QD-cavity coupling strength $g/\kappa = 2$ and $g^2(0) = 0.004$ [17]. The destructive quantum interference effect can be effective to achieve photon antibunching with lower photon nonlinearity. A new mechanism called unconventional photon blockade (UPB), originated from the destructive quantum interference effect in the nonlinear photonic molecule, was proposed to achieve strong photon antibunching with weak nonlinearity in the cavities [18-19]. A number of studies have been implemented to generate antibunching photons with weak Kerr-nonlinearity of the photon based on two coupled photonic cavities [20-23]. Gerace D *et al.* proposed doubly resonant microcavities to achieve UPB with $g^2(0) \approx 10^{-2}$ under the condition that the tunneling strength between two cavities is $J/\kappa \approx 19.45$ [19]. Xu *et al.* proposed a nonlinear photonic molecule with both the two cavity modes being driven coherently to achieve strong photon antibunching with tunneling strength between two cavities $J/\kappa = 10$ [20]. Despite that the photon antibunching is achieved with modest tunneling strength in the cavities, the intracavity photon number which is a key factor for ideal single photon sources with a large cavity output [24] is still low. Lower QD-cavity coupling strength and tunneling strength which avoids the fabrication challenges for preparing microcavities with high quality factors, along with large intracavity photon number are still in need to create practical single photon sources.

In this paper, we propose a scheme to lower the photon nonlinearity requirement for photon antibunching utilizing UPB by means of a photonic molecule. Our system consists of two coupled cavities, and one of which contains a QD. By applying a continuous wave (CW) laser to the system, a strong photon antibunching can be realized by UPB with modest QD-cavity coupling strength. More importantly, when the system is driven by two CW lasers, the need for strong photon antibunching with a large intracavity photon number can be met with much lower cavity tunneling strength compared to single laser driving condition by optimizing the relative phase between the two driving laser fields. Meanwhile, the system can maintain strong photon antibunching within a large range of parameters with the optimum relative phase between two driving laser fields. Thus, our scheme can be used as a tunable single photon source, which is more feasible in experiments.

## 2. Theoretical model

We consider a photonic molecule consisting of two coupled single mode cavities A and B with cavity frequencies $\omega_a$ and $\omega_b$ respectively, which can be achieved experimentally by whispering-gallery-mode optical resonators [25-29], and one of the cavities contains a QD (see the inset in Fig.1). The cavity B is coupled with a two-level QD and cavity A with coupling strength $g$ and tunneling strength $J$ respectively, but there is no interaction between cavity A and the QD. The two cavities are driven by CW laser fields respectively with laser frequency $\omega_L$ and $\omega_d$. Using the rotating wave approximation, the Hamiltonian for the system can be described by (with $\hbar=1$)

$$H = \omega_a a^\dagger a + \omega_b b^\dagger b + \omega_\sigma \sigma^\dagger \sigma + g(\sigma^\dagger b + b^\dagger \sigma) + J(a^\dagger b + b^\dagger a)$$
$$+ E_a(a^\dagger e^{-i\omega_L t - i\theta} + a e^{i\omega_L t + i\theta}) + E_b(b^\dagger e^{-i\omega_d t} + b e^{i\omega_d t}), \quad (1)$$

where $a$ ($b$) and $a^\dagger$ ($b^\dagger$) are the annihilation and creation operators of cavity mode $a$ ($b$) for cavity A (B); $\sigma$ and $\sigma^\dagger$ are the lowering and raising operators for QD. We assume that the QD with the jump frequency $\omega_\sigma$ is resonant to cavity mode $b$, i.e. $\omega_b = \omega_\sigma$; $E_a$ and $E_b$ are the pumping strength of the two driving laser fields; $\theta$ is the relative phase between the two driving laser fields. For simplicity, we utilize the unitary transformation U,

$$U(t) = \exp(-i\omega_L a^\dagger a t - i\omega_d b^\dagger b t - i\omega_d \sigma^\dagger \sigma t), \quad (2)$$

Thus, the interaction Hamiltonian of the coupled system will be time-independent and can be rewritten as

$$H_{eff} = U^\dagger H U - iU^\dagger \frac{\partial U}{\partial t}$$
$$= \Delta_a a^\dagger a + \Delta_b b^\dagger b + \Delta_b \sigma^\dagger \sigma + g(\sigma^\dagger b + b^\dagger \sigma) + J(a^\dagger b + b^\dagger a)$$
$$+ E_a(a^\dagger e^{-i\theta} + a e^{i\theta}) + E_b(b^\dagger + b), \quad (3)$$

where $\Delta_a = \omega_a - \omega_L$, $\Delta_b = \omega_b - \omega_d = \omega_\sigma - \omega_d$. In the following, we will consider the dissipation of the system. The cavity decay rates for cavity A and B are $\kappa_a$ and $\kappa_b$. Although the best combination of detunings and dissipations will lead to further optimal photon antibunching [30], for specific physical picture and brief results, here we assume $\kappa_a = \kappa_b = \kappa$ and $\Delta_a = \Delta_b = \Delta$. The dynamics of the system with dissipation follows the master equation $\dot{\rho} = -i[H, \rho] + L\rho$, where $L$ is the Lindblad superoperator which represents the incoherent loss of the system [31]. Under the low temperature condition, we neglect the influence of phonon, $L\rho$ is given by [16]

$$L\rho = \frac{\kappa}{2} L(a)\rho + \frac{\kappa}{2} L(b)\rho + \frac{\gamma}{2} L(\sigma)\rho, \quad (4)$$

with a definition of $L(x)\rho = 2x\rho x^\dagger - x^\dagger x \rho - \rho x^\dagger x$. Here, $\gamma$ is the spontaneous emission rate for QD. Under the weak driving condition, we can expand the wave function on a Fock-state basis as [18]

$$|\Psi\rangle = \sum_{m=0,n=0}^{\infty} C_{m,n,g}|m,n,g\rangle + \sum_{m=0,n=1}^{\infty} C_{m,n-1,e}|m,n-1,e\rangle, \quad (5)$$

Here, $|m,n,x\rangle$ represents the Fock state with m photons in mode $a$ of cavity A, n photons in mode $b$ of cavity B and the

QD in the ground state (x=g) or excited state (x=e). $C_{m,n,x}^2$ represents the probability of eigenstate $|m,n,x\rangle$. For the UPB case, we just need to truncate the Fock state basis to the two-photon manifold. The wave function can be expanded as

$$|\Psi\rangle = C_{0,0,g}|0,0,g\rangle + C_{1,0,g}|1,0,g\rangle + C_{0,1,g}|0,1,g\rangle + C_{0,0,e}|0,0,e\rangle + C_{2,0,g}|2,0,g\rangle$$
$$+C_{1,1,g}|1,1,g\rangle + C_{0,2,g}|0,2,g\rangle + C_{1,0,e}|1,0,e\rangle + C_{0,1,e}|0,1,e\rangle. \quad (6)$$

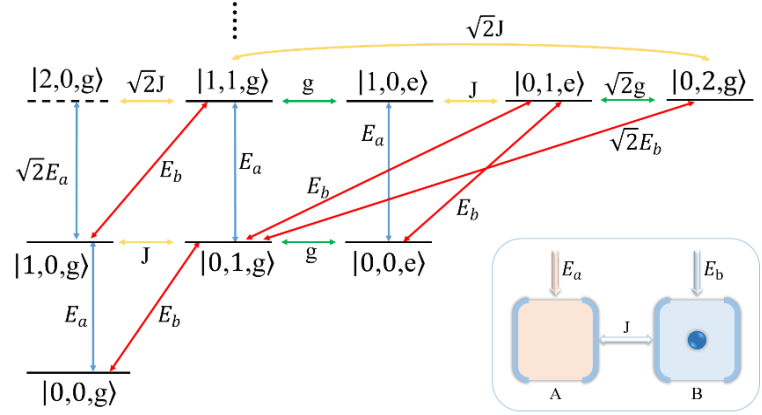

Fig.1 Transition paths for the destructive quantum interference model. The inset shows the scheme of two coupled cavities containing a QD driven by two continuous-wave fields.

To suppress the two photon excitation, the condition $C_{2,0,g}^2 = 0$ is necessary. In this limit, all higher photon excitations with $m \geq 2$ are eliminated, leading to only one excited photon in cavity A. This UPB mechanism is different from the photon blockade in strong coupling regime where the higher photon excitations are far off-resonance owing to anharmonicity of the energy spectrum. The quantum interference effect with different transition paths as shown in Fig.1 is responsible for UPB mechanism.

## 3. Numerical Results

### 3.1. Single laser driving

Utilizing the quantum interference mechanism, we can obtain strong photon antibunching both by applying single laser field or two laser fields on our photonic molecule scheme. Based on this theory, we simulate the dynamics of the system and numerically calculate the photon statistics with the quantum optics toolbox [32]. By solving the time-independent master equation within a truncated Fock space, we can calculate the zero-delay second-order correlation function defined as $g^2(0) = \langle a^\dagger a^\dagger aa\rangle/\langle a^\dagger a\rangle^2$ and the expectation of intracavity photon number to estimate the photon statistics of the system.

First, we consider the situation where only one laser field is applied to the system as we set $E_b = 0$ and $\theta = 0$. Following the transition from $|0,0,g\rangle$ to $|1,0,g\rangle$ excited by the driving field, the interference can happen between three paths, the direct transition $|1,0,g\rangle \xrightarrow{\sqrt{2}E_a} |2,0,g\rangle$, the transition $|1,0,g\rangle \xrightarrow{J} |0,1,g\rangle \xrightarrow{E_a} |1,1,g\rangle \xrightarrow{\sqrt{2}J} |2,0,g\rangle$ and the transition $|1,0,g\rangle \xrightarrow{J} |0,1,g\rangle \xrightarrow{g} |0,0,e\rangle \xrightarrow{E_a} |1,0,e\rangle \xrightarrow{g} |1,1,g\rangle \xrightarrow{\sqrt{2}J} |2,0,g\rangle$. Based on this theory, we calculate the photon statistics in cavity A.

In Fig.2, we show the zero-delay second-order correlation function $g^2(0)$ and photon number as a function of the detuning $\Delta/\kappa$ with different tunneling strength $J$ respectively. As we can see in Fig.2 (a), the photon antibunching becomes more significant with increasing $J$ and the curve dip of the $g^2(0)$ moves from $\Delta = \pm g$ to $\Delta = 0$. When $J = \kappa$, the

$g^2(0)$ curve exhibits J-C like property with two dips at $\Delta = \pm g$. This is because the photons in cavity A induced by laser pump are converted to photons in cavity B with the tunneling strength $J$, which introduces excitation to cavity B that leads to a J-C like model of cavity B. But at the mean time, the cavity photon number is restrained as shown in Fig.2 (b) because the cavity-light detuning is 0. However, much stronger photon antibunching can be obtained at $\Delta = 0$ where the two photon excitaion is extremely restricted due to the quantum interference mechanism. And also, since the driving field is resonant with the eigen frequency of the cavity, the intracavity photon number can maitain large at $\Delta = 0$. When $J = 3\kappa$, we can get strong photon antibunching at $\Delta = 0$ with $g^2(0) \approx 4 \times 10^{-4}$ and intracavity photon number N=0.006. However, the increase of tunneling strength $J$ also reduces single photon emission efficiency due to the photon tunneling between cavities which converts photons of mode *a* to photons of mode *b*. What is more, with the parameters combination, the system can maintain photon antibunching with cavity-light detuning from -0.4$\kappa$ to 0.4$\kappa$ which gives the external driving laser field a large detuning frequency range.

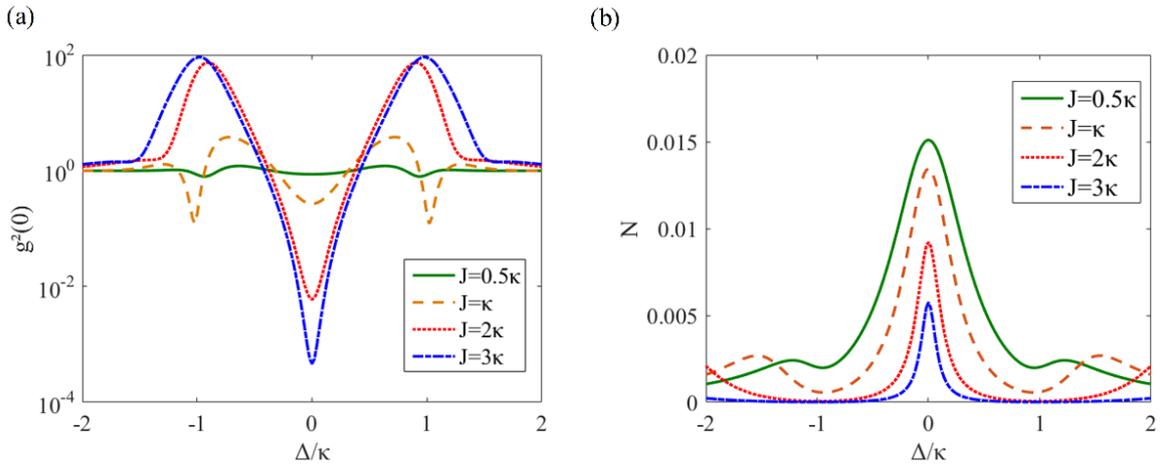

Fig.2 (a) The zero-delay second-order correlation function and (b) the intracavity photon number as functions of cavity-light detuning for different tunneling strengths. Other parameters are $E_a/2\pi$=1 GHz, $\kappa/2\pi$=16 GHz, $g = \kappa$, $\gamma/2\pi$=1 GHz.

To further investigate the optimal photon antibunching, we calculate $g^2(0)$ as a function of QD-cavity coupling strength $g$ with different tunneling strength $J$ in Fig.3 (a). As illustrated in Fig.3 (a), we can find the optimal coupling strength $g$ with different $J$ for strong photon antibunching. When $J = 3\kappa$, we find the optimal photon antibunching can be achieved with $g^2(0) \approx 10^{-4}$ at g = 1.1$\kappa$. Moreover, there is another extremum at g = 2.7$\kappa$ for $J = 3\kappa$. As the tunneling strength increases, more photons of mode *a* convert to photons of mode *b* which introduces excitation to pump the QD-cavity system in cavity B. Thus under the strong coupling regime with g = 2.7$\kappa$, the QD-cavity system can generate antibunching photons due to the nonlinearity of unequal level spacing like J-C model. These antibunching photons in cavity B convert to photons of mode *a* in cavity A through tunneling effect, which leads to another dip of $g^2(0)$ for $J = 3\kappa$. And also, strong photon antibunching can be maintained within a large parameter variation of $g$ which leads to more flexibility for cavity fabrication. Under the optimal photon antibunching condition, we calculate the $g^2(0)$ and intracavity photon number as functions of driving strength as shown in Fig.3 (b). The $g^2(0)$ increases with the augment of laser driving strength, because the increasing driving strength enhances the probability of multiple-photon excitation. Additionally, the intracavity photon number grows very fast while the $g^2(0)$ changes gently with the driving strength ranging from 0.01$\kappa$ to 0.4$\kappa$. As we can see in Fig.3 (b), the domain left to the vertical green dash line represents the situation where photon antibunching is achieved. Thus, according to this, we can get a balance between large photon number and strong photon antibunching. For example, when $E = 0.4\kappa$, the mean intracavity photon number is 0.05 and the $g^2(0)$ is around 0.002. As result, under the single laser driving regime, we can achieve strong antibunching with modest QD-coupling strength.

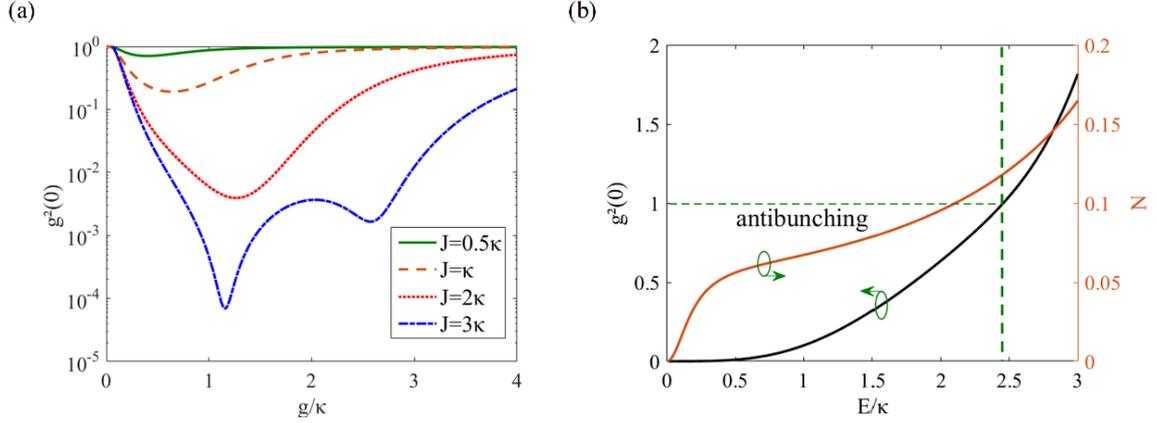

Fig.3 (a) The zero-delay second-order correlation function as a function of the QD-cavity coupling strength for different tunneling strengths with $E_a/2\pi$=1 GHz. (b) The zero-delay second-order correlation function and intracavity photon number as function of driving strength with $g$=1.1$\kappa$, $J$ =3$\kappa$. Other parameters are $\kappa/2\pi$=16 GHz, $\gamma/2\pi$=1 GHz, $\Delta = 0$.

### 3.2. Double laser driving

As we can see in Fig.3 (a), the photon antibunching is not strong even with an optimum QD-cavity coupling strength when $J=\kappa$. We wonder if we can also lower the requirement for tunneling strength between two cavities so that we don't need to put the two cavities too close. And also for practical implementation, tunable single photon sources where the optimal conditions for strong photon antibunching are related to some controllable parameters are in demand. Thus, we consider the situation where two cavities are both driven by laser field respectively. The driving strength and the relative phase between the two laser fields can be considered as tunable parameters for obtaining single photon source. We assume that the two laser fields pump the two cavities simultaneously with the same driving strength $E = E_a = E_b$ for simplicity, and the relative phase between two driving field is $\theta$. Following the transitions from $|0,0,g\rangle$ to $|1,0,g\rangle$ and $|0,1,g\rangle$ excited by the two driving laser fields $E_a$ and $E_b$, the UPB can be achieved by destructive quantum interference between multiple transition paths shown in Fig. 1, mainly the transition $|1,0,g\rangle \xrightarrow{\sqrt{2}E_a} |2,0,g\rangle$, $|1,0,g\rangle \xrightarrow{E_b} |1,1,g\rangle \xrightarrow{\sqrt{2}J} |2,0,g\rangle$, $|0,1,g\rangle \xrightarrow{E_a} |1,1,g\rangle \xrightarrow{\sqrt{2}J} |2,0,g\rangle$, $|1,0,g\rangle \xrightarrow{J} |0,1,g\rangle \xrightarrow{E_a} |1,1,g\rangle \xrightarrow{\sqrt{2}J} |2,0,g\rangle$, and transition paths excited from $|0,1,g\rangle$ and $|0,0,e\rangle$ by $E_b$ will all contribute to the former transitions. Based on this understanding, we calculate the photon statistics in cavity A.

In this regime, the relative phase $\theta$ between two driving field is the main reason for the quantum interference induced UPB. The zero-delay second-order correlation function $g^2(0)$ as a function of the cavity-light detuning and relative phase is shown in Fig.4 (a). We can find that the photon statistics is strongly related to relative phase. And also, the $g^2(0)$ exhibits an antisymmetry like property due to the different transition paths excited by driving fields with different cavity-light detuning and relative phase. Thus, we can change the photon statistics just by adjusting the frequency of external driving fields.

Then, we calculate the intracavity photon number as a function of the cavity-light detuning with different relative phases as shown in Fig.4 (b). We find that the intracavity photon number reaches a peak around $\Delta = 0$ for different relative phases. For an ideal single photon source with a large cavity output, large intracavity photon number is a key factor. In order to achieve strong photon antibunching and maintain a large intracavity photon number at the same time, we choose the optimal value with $\Delta = 0$ and $\theta$=1.5$\pi$.

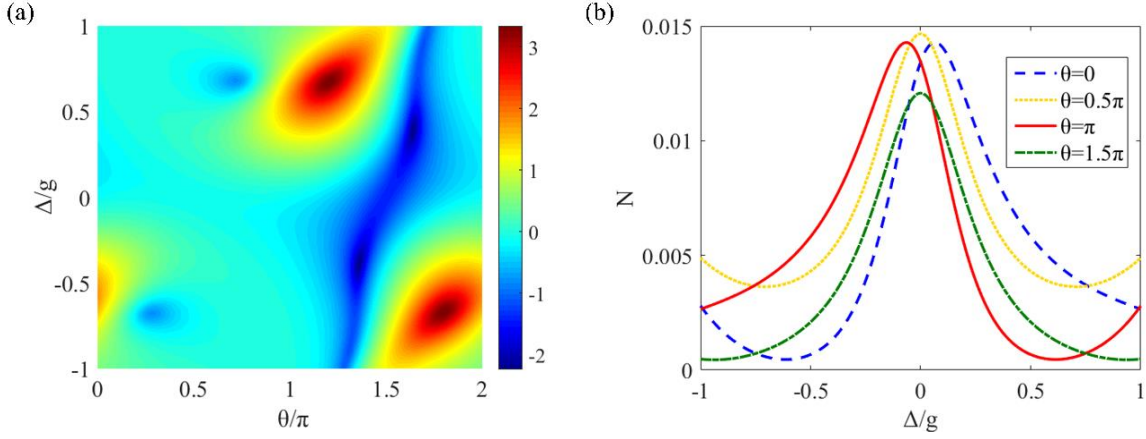

Fig.4 (a) The zero-delay second-order correlation function in logarithmic scale as a function of the cavity-light detuning $\Delta$ and relative phase $\theta$ and (b) the intracavity photon number $N$ as a function of cavity-light detuning $\Delta$ for different relative phases $\theta$ with $\kappa/2\pi$=16 GHz, $E/2\pi$=1 GHz, $\gamma/2\pi$=1 GHz, $J=\kappa$, $g=\kappa$.

For further estimation of the system, we set $\Delta = 0$ and calculate the $g^2(0)$ as a function of relative phase $\theta$ with different tunneling strength $J$ as shown in Fig.5 (a). For different tunneling strength $J$, photon antibunching can maintain for different relative phase as $g^2(0) < 0$ and we can achieve optimal photon antibunching at relative phase $\theta$=-0.5π and $\theta$=1.5π. Especially for $J=\kappa$, we can achieve strong photon antibunching with $g^2(0) = 0.035$ at $\theta$=1.5π. For $\theta$=1.5π, the perfect destructive quantum interference is achieved to eliminate $|2,0,g\rangle$ in Fig.1. To find the optimal $J$ for strong photon antibunching, we calculate the $g^2(0)$ as a function of tunneling strength $J$ with $\theta$=1.5π in Fig.5 (b). We find an obvious dip of $g^2(0)$ at $J$=0.9$\kappa$ with $g^2(0) = 0.002$ indicating an optimal tunneling strength value for UPB. And also, we find that photon antibunching is suppressed when $J > 0.9\kappa$. The reason is that although the tunneling between two cavities can provide extra transition paths compared to a QD-cavity system, but the interaction between two cavities also converts photons of mode *a* to photons of mode *b* which decreases the single photons generated in cavity A. So when the tunneling strength is too large, the single photon generation will be suppressed instead.

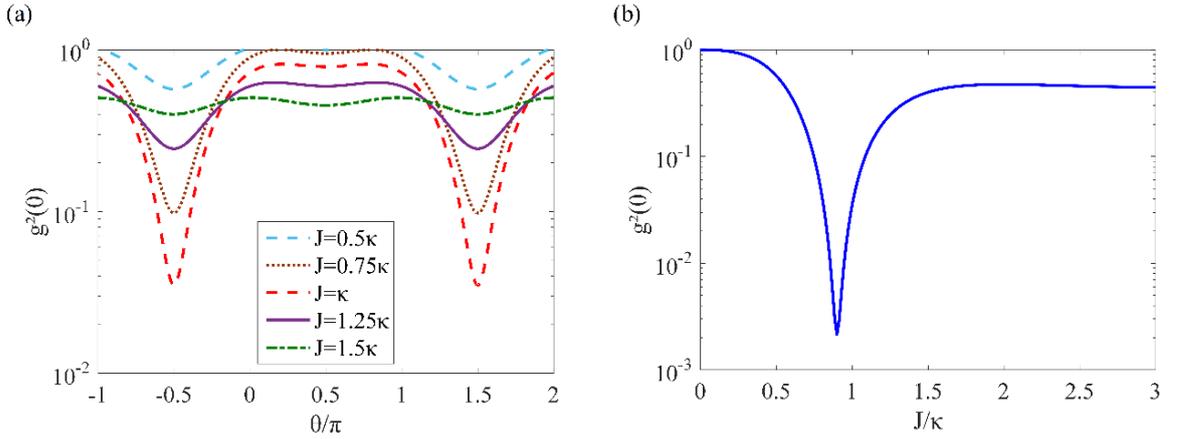

Fig.5 (a) The zero-delay second-order correlation function as a function of the relative phase $\theta$ with different tunneling strengths. (b) The zero-delay second-order correlation function as a function of tunneling strength with $\theta$=1.5π.

Then, we investigate the robustness of the double driving scheme for stucture parameters. The second-order correlation function and the intracavity photon number as functions of QD-cavity coupling strength g and tunneling strength $J$ are illustrated in Fig.6 (a) and (b). We find our scheme can maintain strong photon antibunching for a large parameters variation as the dark blue area shown in Fig.6 (a) and strong photon bunching as the yellow area shown in Fig.6 (a). In Fig.6 (b), a large intracavity photon number can be obtained under the strong photon antibunching condition while much lower intracavity photon number is observed for photon bunching condition, which indicates that the two-photon excitaion is suppressed under the double laser driving scheme by UPB. And also, the $g^2(0)$ keeps quasi-linearly relation with $g$ and $J$ under antibunching condition as the green dash line shown in Fig.6 (a). Especially, we can achieve strong photon antibunching with QD-cavity coupling strength and tunneling strength as low as $g=0.18\kappa$ and $J=0.48\kappa$. Thus, this regime can be used as a tunable single photon source that lowers the requirement for high-quality microcavities.

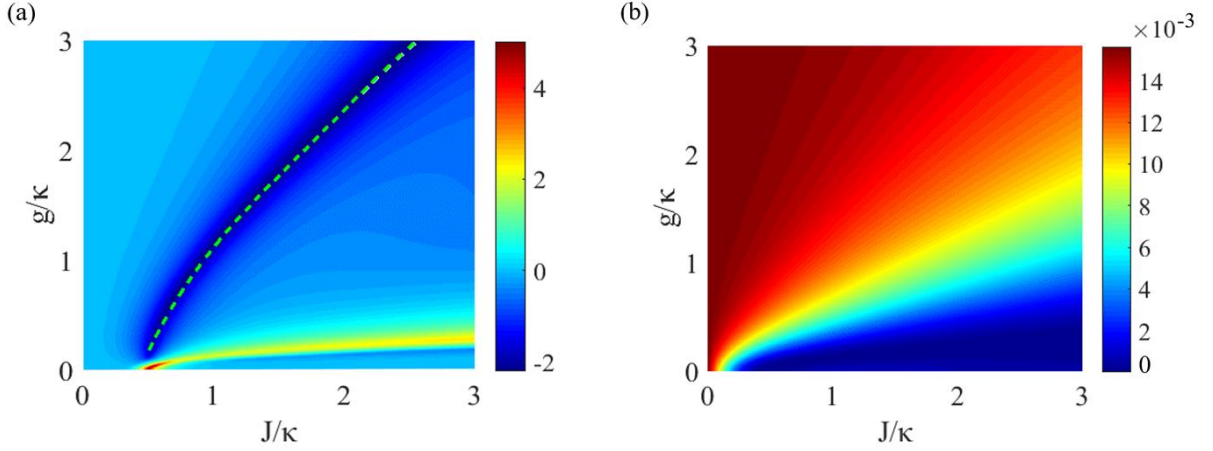

Fig.6 (a) The zero-delay second-order correlation function in logarithmic scale and (b) the intracavity photon number as functions of the QD-cavity coupling strength and tunneling strength. Other parameters used for the calculation above are $\theta=1.5\pi$, $\kappa/2\pi=16$ GHz, $E/2\pi=1$ GHz, $\gamma/2\pi=1$ GHz. The green dash line represents the lowest $g^2(0)$ with the parameters combination.

In Fig.7 (a), we calculate the secong-order correlation function and the intracavity photon number as functions of the strength of driving fields. As we can see, the $g^2(0)$ and intracavity photon number are both increasing with the enhancement of the driving field. When the drving strength is as low as $E/2\pi=0.5$ GHz, the $g^2(0)$ is around $10^{-3}$ with a relatively low intracavity photon number. On the other hand, if we want large intracavity photon number such as N=0.1 with the driving strength $E/2\pi=3.5$ GHz, the system can still maintain photon antibunching with $g^2(0) \approx 0.09$.

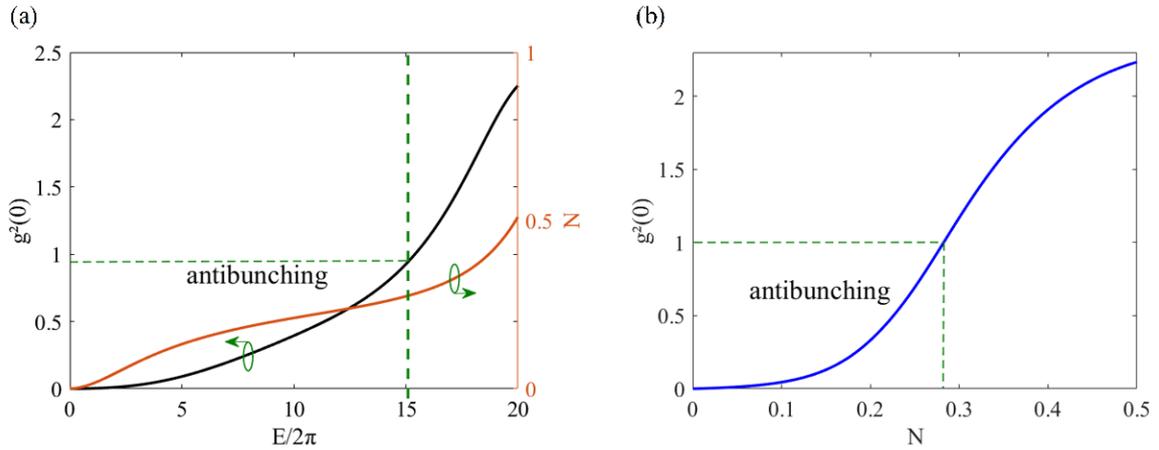

Fig.7 (a) The zero-delay second-order correlation function and the intracavity photon number $N$ as functions of the driving strength, and (b) the zero-delay second-order correlation function as a function of the intracavity photon number $N$ for changeable driving strengths with the parameters $\kappa/2\pi=16$ GHz, $\gamma/2\pi=1$ GHz, $g=\kappa$, $J=0.9\kappa$, $\theta=1.5\pi$.

For practical implementation, we need to take both photon antibunching and photon number into consideration. In order to get a balance, we calculate the $g^2(0)$ as a function of intracavity photon number as shown in Fig.7 (b). The $g^2(0)$ increases with the enhancement of the intracavity photon number, and the photon antibunching can be sustained with the intracavity photon number around 0.28 at most. This can be used as a guidance to choose the most appropriate parameter values of the scheme for single photon generation in experiments.

## 4. Conclusions

In summary, we proposed a scheme of two photonic cavities, one of which contains a QD, to achieve strong photon antibunching with lower QD-cavity coupling strength and tunneling strength between two cavities. The system can achieve UPB by utilizing destructive quantum interference to suppress the two-photon excitation. Strong photon antibunching ( $g^2(0) \approx 10^{-4}$ ) can be obtained with modest QD-cavity coupling strength when the system is driven by single laser field. To further lower the QD-cavity coupling strength and tunneling strength and make the system tunable, we apply two laser fields to the system. By optimizing the relative phase between the two driving laser fields, the strong photon antibunching ( $g^2(0) \approx 10^{-3}$ ) can be achieved with a relatively large intracavity photon number under weak QD-cavity coupling strength and tunneling strength. Under the optimal phase condition, the system also shows a good robustness of maintaining strong photon antibunching within a large parameter variation. Thus, our proposed scheme, which can be easily realized experimentally, can serve as a tunable single photon source. The system may have potential applications in quantum information processing.

## Acknowledgements

This work was supported by the National Natural Science Foundation of China (Grants No.61372037and No.61401035), the Fund of State Key Laboratory of Information Photonics and Optical Communications (Beijing University of Posts and Telecommunications), P. R. China (Grant No. IPOC2015ZC05).